\documentclass[useAMS,usenatbib,usegraphicx]{mn2e}
\onecolumn

\usepackage[T2A]{fontenc}
\usepackage[english]{babel}
\usepackage[dvips]{graphicx}
\usepackage{mathtext}

\title[Perturbations of ionization fractions]
{Perturbations of ionization fractions  at the cosmological recombination epoch}

\author[B. Novosyadlyj]
{ B. Novosyadlyj\thanks{E-mail: novos@astro.franko.lviv.ua} \\
Astronomical Observatory of the Ivan Franko National University of Lviv, Kyryla i Methodia str., 8, Lviv, 79005, Ukraine}  
\begin{document}

\date{Accepted 2006 May 17 . Received 2006 March 22}

\pagerange{\pageref{firstpage}--\pageref{lastpage}} \pubyear{2006}

\maketitle

\label{firstpage}

\begin{abstract}
 
 A development of perturbations of  number densities  of ions and electrons during recombination epoch is analysed. 
The equations for relative perturbations of ionization fractions were derived from the system of equations 
for  accurate computation of the ionization history of the early Universe given by \citet{seager1999,seager2000}. 
It is shown that strong dependence of ionization and recombination rates on the density and temperature of  plasma 
provides the significant deviations of amplitudes of  ionization fractions relative perturbations from ones of baryon matter 
density adiabatic perturbations. Such deviations are most prominent for cosmological adiabatic perturbations of  scales larger 
than sound horizon at recombination epoch. The amplitudes of relative perturbations of number densities of electrons and protons 
at last scattering surface exceed  by factor of $\simeq$5  the amplitude of relative perturbation of baryons total number density, 
for helium ions this ratio reaches the value of $\simeq$18. For subhorizon cosmological perturbations these ratios appear to be 
essentially lesser and depend on oscillation phase at the moment of decoupling. These perturbations of number densities of ions 
and electrons at recombination epoch do not contribute to the intrinsic plasma temperature fluctuations but cause the ''corrugation'' 
of last scattering surface in optical depth, $\delta z_{dec}/(z_{dec}+1)\approx -\delta_b/3$, at scales larger than sound horizon. 
It may result into noticeable changes of precalculated values of CMB polarization pattern at several degrees angular scales.

\end{abstract}


 

\section*{Introduction}

Cosmic microwave background (CMB) radiation coming from recombination epoch has become one of the most powerful observational probes for cosmological models of our Universe and formation of its large-scale structure. Indeed, the full-sky maps of  cosmic microwave temperature fluctuations  obtained by Wilkinson Microwave Anisotropy Probe (WMAP)  during first year of observations  have given a possibility to determine the cosmological parameters with high accuracy, $\sim2\%$ \citep{bennett2003,verde2003,spergel2003}. The current data from three year of  WMAP  observations \citep{spergel2006} and, especially, expected from future mission Planck, improve precision of the CMB power spectrum determination to the level of accuracy of  numerical precalculations done by the most advanced codes in the framework of specific model. For example, CMBfast code by \citet{cmbfast96,cmbfast99} has intrinsic accuracy $\simeq 1\%$.  An adequate calculation of the recombination process is crucial for modelling the power spectrum of CMB temperature fluctuations and polarization. 

The first analyses of recombination kinetics were carried out by  \citet{zeldovich1968} and \citet{peebles1968}  in 1967. In  subsequent papers     \citep{matsuda1971,zabotin1982,lyubarsky1983,jones1985,krolik1990,rubicki1993}
the main processes have been studied  using  the 3-level approximation of hydrogen and helium atoms. An accuracy of few percents has been achieved.  The most complete analysis of cosmological recombination processes with taking into account the multi-level structure of hydrogen 
and helium atoms ($\simeq300$ levels) and  non-equilibrium ionization-recombination kinetics has been performed  by Seager, Sasselov \& Scott (2000). 
 Also all known plasma thermal processes were taken into account therein. These authors have  provided cosmological community with software  
 RECFAST  \citep{seager1999} which ensures an accuracy of calculation of number density of electrons $\sim 1\%$. This code was used by number 
 of authors to calculate the transfer function of density perturbations and power spectrum of  CMB temperature fluctuations and polarization, 
 in particular, by \citet{cmbfast96,cmbfast99} for publicly available software CMBFAST. However, the researches aimed on improving the 
 calculation of  recombination and decoupling of the thermal radiation from baryon plasma are still going on (see recent papers by 
 \citet{dubrovich2005,chluba2006,kholupenko2005,wong2005} and citing therein).


In this paper the evolution of number density perturbations of hydrogen and helium ions and electrons  in the field of  cosmological matter density perturbations is studied. The cosmological adiabatic perturbations closely  associate the density and temperature variations in baryon-radiation component, leading to  the corresponding variations of photorecombination and photoionization rates, which in turn can cause  appreciable  deviations of relative perturbation amplitudes of number density  of ions and electrons from corresponding amplitude for total number density of baryon nuclei. The effective 3-level models of hydrogen and helium atoms by  \citet{seager1999}  and their software RECFAST  were used as a basis. Here all calculations  were carried out for $z>100$ when recombination and dissociation processes of hydrogen negative ions ${\rm H^-}$ and molecules  ${\rm H_2}$ and
 ${\rm H_2^+}$ can be neglected due to their insignificance.

In the first section the basic equations for  hydrogen and helium recombination in homogeneous expanding Universe are presented along with their numerical solutions for $\Lambda$CDM model. The definitions of relative 
perturbations of  number densities of ions and electrons, their properties, equations for  their evolution and  results of  integration for the stationary mass density and temperature initial perturbations are given in the second section. The third section is devoted to the analysis of perturbations of electron number density  within the adiabatic mass density perturbations of different scales in $\Lambda$CDM model. An estimation of contribution of this effect into the ''corrugation'' of last scattering surface and CMB temperature fluctuations is carried out in the 4th section.

\section{Cosmological recombination of  hydrogen and helium atoms: definitions, equations, results}

Let us introduce the following definitions:       
$n_{\rm HI}$ and $n_{\rm HII}$ denote the number densities of neutral  and ionized hydrogen atoms respectively, 
$n_{\rm HeI}$, $n_{\rm HeII}$ and $n_{\rm HeIII}$ -- number densities of neutral, singly and double ionized helium atoms,
$n_{\rm e}=n_{\rm HII}+n_{\rm HeII}+2n_{\rm HeIII}$ -- number density of free electrons,
$n_{\rm H}=n_{\rm HI}+n_{\rm HII}$ -- total number density of  hydrogen nuclei,  $n_{\rm He}=n_{\rm HeI}+n_{\rm HeII}+n_{\rm HeIII}$ -- total number density of helium nuclei. It is conveniently to use the relative number densities (ionization fractions):  $x_{\rm HI}\equiv n_{\rm HI}/n_{\rm H}$ -- relative abundance of neutral hydrogen, $x_{\rm HII}\equiv n_{\rm HII}/n_{\rm H}$ -- the same for ionized hydrogen,  $x_{\rm HeI}\equiv n_{\rm HeI}/n_{\rm He}$, $x_{\rm HeII}\equiv n_{\rm HeII}/n_{\rm He}$, 
$x_{\rm HeIII}\equiv n_{\rm HeIII}/n_{\rm He}$ -- relative abundances of neutral helium,  singly and double ionized helium atoms, $x_{\rm e}\equiv n_{\rm e}/n_{\rm H}$ --  relative number density of electrons. The ratio of total number densities of helium and hydrogen nuclei we define as $f_{\rm He}\equiv n_{\rm He}/n_{\rm H}$, which can be expressed via mass fraction of primordial helium $Y_P$, so that $f_{\rm He}=Y_P/4(1-Y_P)$ (further we assume $Y_P=0.24$ from \citet{schramm1998}).
This quantities obey evident relationships: 
$x_{\rm e}=x_{\rm HII}+f_{\rm He}x_{\rm HeII}+2f_{\rm He}x_{\rm HeIII}$, $x_{\rm HI}+x_{\rm HII}=1$, 
$x_{\rm HeI}+x_{\rm HeII}+x_{\rm HeIII}=1$. All following formulae and relations are presented for these relative number densities
of atoms, ions and electrons.

As it follows from above cited papers and, in particular, from refined numerical calculations by \citet{seager1999,seager2000}  at early stages of  universe evolution ($z>10^4$) all hydrogen and helium atoms were ionized completely by thermal  photons, so, then $x_{\rm HII}=1$, $x_{\rm HI}=0$, $x_{\rm HeIII}=1$, $x_{\rm HeI}=x_{\rm HeII}=0$  and $x_{\rm e}=1+2f_{\rm He}$. This is a consequence of  high number density of thermal  high-energy photons capable to rend electrons from all atoms and ions.  In the expanding Universe  the energy of each photon and  temperature of radiation decrease $\propto a^{-1}$, radiation energy density $\propto a^{-4}$, the number and mass
density of baryons and dark matter $\propto a^{-3}$, where $a$ is the scale factor which is associated with redshift $z$ by simple relation $a=(z+1)^{-1}$.  Already at $z\sim 8000$ thermal photons with energies higher than potentials of ionization 
of ${\rm HeII}$ from ground and second levels reside in the short-wave tail of Planck function and their number density becomes too low to keep all helium in the ionization stage of ${\rm HeIII}$. It begins to recombine and at $z\sim 7000$ the $\rm HeII$ ions  appear. At this moment the time-scale of Thomson scattering ($t_{\rm T} \simeq  3m_ec(1+x_{\rm e}+f_{\rm He})/(8\sigma_{\rm T}a_{\rm R} T_{\rm R}^4x_{\rm e})$,   
the hydrogen recombination time-scale ($t_{\rm HI}\simeq 1/n_{\rm e}\alpha_{\rm HI}$) and the helium one ($t_{\rm HeII}\simeq 1/n_{\rm e}\alpha_{\rm HeII}$) appear to be essentially lower in comparison with the time-scale of Universe expansion  ($t_{\rm Hubble}\simeq 2/3H_0(1+z)^{3/2}$).  Thus, matter temperature (electronic and ionic one) $T_{\rm m}$ equals to CMB temperature,  $T_{\rm R}$. Recombination of $\rm HeII$ occurs in the conditions of local thermodynamic equilibrium (LTE). In the expressions for time-scales  $m_e$ denotes the mass of electron, $c$ is the light speed, $\sigma_{\rm T}$ is the effective  cross-section of Thomson scattering, $a_{\rm R}$ is the radiation constant, $\alpha_i$ is the effective coefficients of recombination to the ground states of hydrogen atoms HI and singly ionized helium atoms, HeII. An ionization fraction of 
helium, $x_{\rm HeIII}$, is described by Saha equation: 
\begin{equation}
\label{SahaHeIII}
{x_{\rm e}x_{\rm HeIII}\over x_{\rm HeII}} =
 {(2\pi m_{\rm e} k T_m)^{3/2}\over h^3 n_{\rm H}}
 {\rm e}^{-\chi_{\rm HeII}/kT_m}.
\end{equation}
Whereas at this epoch both hydrogen and helium  atoms are completely ionized  ($x_{\rm HI}=0$, $x_{\rm HII}=1$, 
$x_{\rm HeI}=0$),  $x_{\rm HeII}=1-x_{\rm HeIII}$  that means $x_{\rm e}=1+f_{\rm He}(1+x_{\rm HeIII})$ and 
above equation can be easily solved for $x_{\rm e}$. It  proves that  already at $z\sim 5000$ all helium atoms become  
singly ionized. This holds up to $z\sim 3500$ when $\rm HeI$ begins to recombine.  At this stage  
$t_{\rm Hubble}:t_{\rm T}:t_{\rm HI}:t_{\rm HeI}\simeq 1:0.0000003:0.0003:0.001$ and conditions are close to LTE. Until the part of HeI 
constitutes less than 1\%  of  total  helium content the metastable $2s$ level plays insignificant role  in deviation of radiative recombination rate of ${\rm HeI}$ from LTE one  and ionized fraction $x_{\rm HеII}$ is described yet enough accurately by Saha equation  
\begin{equation}
\label{SahaHeII}
{x_{\rm e}x_{\rm HeII}\over x_{\rm HeI}} =
 4{(2\pi m_{\rm e} k T_m)^{3/2}\over h^3 n_{\rm H}}
 {\rm e}^{-\chi_{\rm He I}/kT_m}.
\end{equation}
Now $x_{\rm HeIII}=0$ and $x_{\rm HeI}= 1-x_{\rm HeII}$. To run an accurate calculation of $x_{\rm HeII}$ we have to obtain the exact value of $x_{\rm e}=x_{\rm HII}+f_{\rm He}x_{\rm HeII}$.  Approximation $x_{\rm HII}=1$ is  already too rough since 0.1\%-decreasing of  $n_{\rm HII}$ due to hydrogen recombination results in comparable change of $n_{\rm e}$ from ${\rm HeI}$ recombination because of the prevalent content of hydrogen ($f_{\rm H}=n_{\rm H}/(n_{\rm H}+n_{\rm He})=0.921$). So, hydrogen recombination must be taken into account too. At this stage it is described by Saha equation:  
\begin{equation}
\label{SahaHII}
{x_{\rm e}x_{\rm HII}\over x_{\rm HI}} =
 {(2\pi m_{\rm e} k T_m)^{3/2}\over h^3 n_{\rm H}}
 {\rm e}^{-\chi_{\rm HI}/kT_m}.
\end{equation}
The system of these two equations can be reduced to cubic algebraic equation for  $x_{\rm e}$ which has one real root:
\begin{equation}
 \label{xe3000}
 x_{\rm e}=2\sqrt{-A/3}\cos{(\alpha/3)}-B/3,
 \end{equation}
where $B=R_{\rm HI}+R_{\rm HeI}$,  $R_{\rm HeI}$ and $R_{\rm HI}$ are right-hand parts of  equations (\ref{SahaHeII}) and (\ref{SahaHII}), $\cos{\alpha}=C/2\sqrt{-A^3/27}$, $A=D-B^2/3$, 
$D=R_{\rm HI}R_{\rm HeI}-R_{\rm HI}-f_{\rm He}R_{\rm HeI}$, $C=2B^3/27-BD/3-E$, 
$E=-R_{\rm HI}R_{\rm HeI}(1-f_{\rm He})$. I have complemented the code RECFAST by this solution 
to achieve correct solution of task formulated above on number density perturbations of ions. However, it does not affect   
the results of calculations for $x$'s noticeably. 

However, such simple description of joint hydrogen-helium recombination loses accuracy very soon, at $z\approx 2800$. Primarily the recombination becomes unequilibrium for HeI because particularity of radiative atomic processes in the expanding cooling Universe.
Both recombination to the ground state and photoionization from it can be omitted because any recombination directly to the ground state will emit 
a photon with energy greater than potential of ionization and it will immediately ionize a neighbouring atom. So, the case B recombination  takes place for
HeII as well as  for HII. Excited atoms of $n\ge 2$ states are ionized  by photons of lower energy, belonging to continua of the second and following series, where the number density of photons is larger  than one of the basic series. Therefore the entire recombination process 
 of ${\rm HeI}$ slows down. The cool radiation field which is very strong causes the ''bottleneck'' effect (for details see \citet{seager2000}) creating 
 the overpopulation of excited states relatively to the Boltzmann distribution. So, the Saha equation does not describe adequately the recombination 
 more and equations of detailed balance must be utilized. In the effective three-level atom  approximation they lead to a single differential equation 
 for the ionization fraction of HeII \citep{seager1999}:

\begin{eqnarray}
\label{eqHeII}
\displaystyle{dx_{\rm HeII}\over \displaystyle dz} =
 {\displaystyle x_{\rm HeII}x_{\rm e} n_{\rm H} \alpha_{\rm HeII}
   - \beta_{\rm HeI} (1-x_{\rm HeII})
   {\rm e}^{-h\nu_{\rm HeI2^1s}/kT_{\rm m}}\over \displaystyle H(z)(1+z)}
 \displaystyle {1 + K_{\rm HeI} \Lambda_{\rm He} n_{\rm H}
  (1-x_{\rm HeII}){\rm e}^{-h\nu_{ps}/kT_{\rm m}}
  \over \displaystyle 1+K_{\rm HeI}
  (\Lambda_{\rm He} + \beta_{\rm HeI}) n_{\rm H} (1-x_{\rm HeII})
  {\rm e}^{-h\nu_{ps}/kT_{\rm m}}}\;\;,
\end{eqnarray}
 where 
\begin{eqnarray}
\label{alphaHeI}
 \alpha_{\rm HeI} =q\left[\sqrt{T_{\rm m}\over T_2}\left(1+\sqrt{T_{\rm m}\over T_2}\right)^{1-p}
 \left(1+\sqrt{T_{\rm m}\over T_1}\right)^{1+p}\right]^{-1}\! 
 \end{eqnarray}
 is effective recombination coefficient ($\mathrm {m^{3}s^{-1}}$) of HeI  \citep{hummer1998}, $\beta_{\rm HeI}$ --  photoionization coefficient,  
 $K_{\rm HeI}\equiv \lambda^3_{\rm HeI2^1p}/[8\pi H(z)]$ -- factor which takes into account the cosmological redshifting of
 HeI  $2^1p-1^1s$  photons. The values for the rest of parameters included into the right-hand part of equation (\ref{alphaHeI}) are listed in
 Table \ref{tab}. 
  Like in the previous case the rate equation for ${\rm HeI}$ recombination  ought to be integrated jointly with equation for HI recombination. 
 Until $n_{\rm HI}$ is still lesser than 1\% of $n_{\rm H}$ the hydrogen ionization fraction, $x_{\rm HII}$,  it can be calculated enough  accurate 
 using the Saha equation ($1600\le z \le 2800$). Optical depth for  Ly$\alpha$ emission increases with  growth of number of hydrogen atoms in the ground state. Diffuse   Ly$\alpha$
 photons, two-photon absorption and collisions as well as cascade recombination from upper levels result in overpopulation of the first excited level.
Instantaneous spontaneous transition $2p-1s$ originates  Ly$\alpha$ photon which is reabsorbed immediately by neighbouring HI atoms in the ground state and distribution of levels populations remains unchanged (case B recombination). The metastable state $2s$ of HI is very important for the recombination kinetics because at these redshifts the probability of two-photon $2s-1s$ transition is much smaller than one of photoionization. Therefore, hydrogen atoms can be ionized from the first excited state by photons of  Balmer continuum, the number density in which exceeds the one of Lyman continuum.
Jointly with the overpopulation of  upper levels relative to a Boltzmann distribution in the strong cool radiation field  (the ''bottleneck'' effect \citet{seager2000})  it leads to non-equilibrium kinetics of recombination.  In this case the equation of detailed balance  must be used to find 
 hydrogen ionization fraction \citep{peebles1968,seager1999}:
 \begin{eqnarray}
\label{eqHII}
\frac{\displaystyle dx_{\rm HII}}{\displaystyle dz} = \frac{\displaystyle  x_{\rm e}x_{\rm HII} n_{\rm H} \alpha_{\rm H}
 - \beta_{\rm H} (1-x_{\rm HII}){\rm e}^{-h\nu_{\rm H2s}/kT_{\rm m}}}{\displaystyle H(z)(1+z)}
  \frac{\displaystyle 1 + K_{\rm H} \Lambda_{\rm H} n_{\rm H}(1-x_{\rm HII})}
    {\displaystyle 1+K_{\rm H} (\Lambda_{\rm H} + \beta_{\rm H})
     n_{\rm H} (1-x_{\rm HII})},
\end{eqnarray}
 where
 \begin{eqnarray}
\label{alphaHI}
\alpha_{\rm H} = F\cdot 10^{-19}at^{b}/(1 + ct^{d}) \;\;\;  \mathrm{m^{3}s^{-1}}
 \end{eqnarray}
  hydrogen recombination coefficient  \citep{pequignot1991}, $t=T_m/10^4$,
  $K_{\rm H}\equiv \lambda^3_{\rm H2p}/[8\pi H(z)]$ -- factor which take into account cosmological redshifting of
Ly$\alpha$  photons. Values for rest of parameters  are presented in Table \ref{tab}. Photoionization coefficients in  
  (\ref{eqHeII}) and  (\ref{eqHII}) are calculated via the case B recombination coefficients in the following way:
 \begin{eqnarray}
\label{beta}
 \beta=\alpha (2\pi m_{\rm e} k T_{\rm m}/h^2)^{3/2}e^{-h\nu_{2s-1s}/kT_{\rm m}}. 
 \end{eqnarray}

The temperature of electrons and ions, $T_{\rm m}$, practically coincides with radiation temperature, $T_{\rm R}$, in the range before $z\sim 800$, since 
until this moment the time-scale of Thomson scattering remains essentially lower than the time-scale of Universe expansion, $t_{\rm T}/t_{\rm Hubble}<10^{-3}$. Therefore, until this moment the rate of temperature decreasing  is governed by adiabatic cooling of radiation ($\gamma=4/3$) caused by Universe expansion:
\begin{equation}
 \frac{dT_{\rm m}}{dz} =\frac{T_{\rm m}}{(1+z)}.
  \end{equation}
After recombination, at  $z<800$, adiabatic cooling of ideal gas ($\gamma=5/3$) begins to dominate over the heating  caused by
Compton effect which is a main process of energy transfer between electrons and photons. 
Cooling of plasma via free-free, free-bound and bound-bound transitions and collisional ionization as well as heating via photoionization and collisional recombination gives  insignificant contribution into the rate of temperature change, it does not  exceed 
the  0.01\% of main processes -- adiabatic cooling and heating by Compton effect  \citep{seager2000}. So, at this epoch the following
equation for the rate of temperature change proves to be enough accurate  \citep{weymann1965,peebles1968,seager1999}:
\begin{equation}
\label{Tm2}
 \frac{dT_{\rm m}}{dz} = \frac{8\sigma_{\rm T}a_{\rm R}
   T_{\rm R}^4}{3H(z)(1+z)m_{\rm e}c}\,
  \frac{x_{\rm e}(T_{\rm m} - T_{\rm R})}{1+f_{\rm He}+x_{\rm e}}
  + \frac{2T_{\rm m}}{(1+z)},
  \end{equation}
The Table \ref{tab} lists the values of all atomic constants and coefficients of approximation formulae seen in equations (\ref{SahaHeIII})-(\ref{Tm2}).

\begin{figure}
\centerline{\includegraphics[width=8cm]{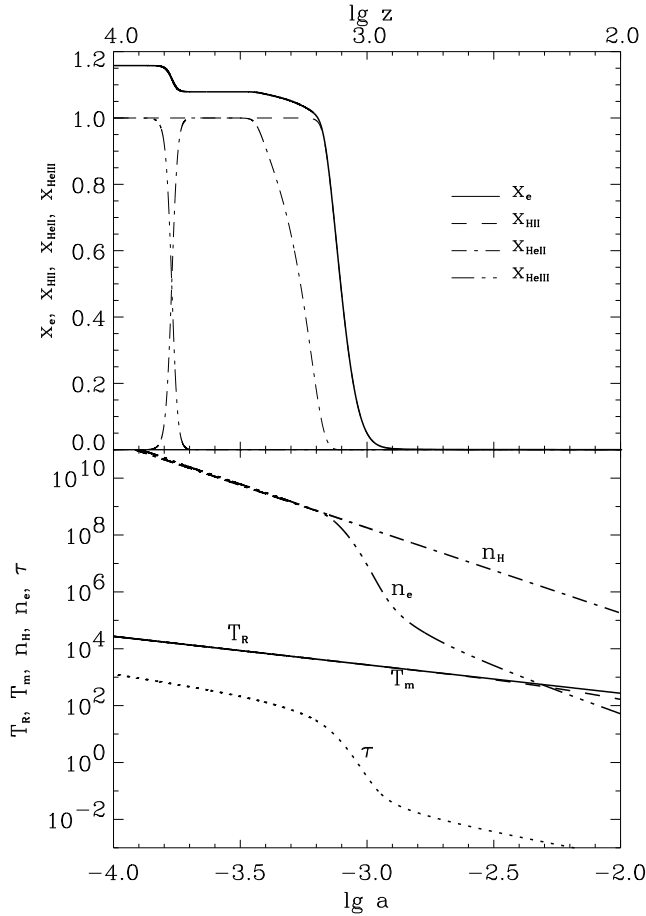}}
\caption{Hydrogen and helium recombination in ${\rm \Lambda CDM}$ model ($\Omega_b=0.05$, $\Omega_{\rm CDM}=0.3$, $\Omega_{\Lambda}=0.65$, $h=0.65$) (top panel). The dependences of  radiation temperature $T_{\rm R}$ (K), matter temperature $T_m$ (K), total hydrogen number density $n_{\rm H}$ (m$^{-3}$), number density of electrons $n_{\rm e}$ (m$^{-3}$) and Thomson scattering optical depth $\tau=\int_0^a {c\sigma_{\rm T} n_{\rm e}H^{-1}d\ln{a}}$  on scale factor $a=(z+1)^{-1}$  (bottom panel).}
\label{rec-Tm}
\end{figure}

The results of calculations of ionization history in the ${\rm \Lambda CDM}$ model of the Universe, performed on the base 
of equations (\ref{SahaHeIII})-(\ref{Tm2}) and  RECFAST code  complemented by solution (\ref{xe3000}), are presented in Fig.\ref{rec-Tm}. 
There are  also shown  the dependences of  radiation temperature $T_{\rm R}$, matter temperature $T_m$, number density of  hydrogen  nuclei $n_{\rm H}$, number density of electrons $n_{\rm e}$ and optical depth $\tau$ due to Thomson scattering by electrons $\tau(z)=\int_0^z {c\sigma_{\rm T} n_{\rm e}(z) H^{-1}(z)(z+1)^{-1}dz}$  on redshift $z$ (top abscissa axis in figure).
In consequence of  rapid expansion of the Universe and non-equilibrium kinetics of recombination-ionisation processes the hydrogen and helium recombinations do not finish with absolutely  neutral medium and some ionized fractions persist at low redshifts that is called the residual ionization. The calculations show that at $z=200$ $x_{\rm e}\approx x_{\rm HII}=6.7\cdot 10^{-4}$ (Fig.\ref{rec-Tm}), $x_{\rm HeII}=9.2\cdot 10^{-10}$ and at $z=0$ $x_{\rm e}\approx x_{\rm HII}=4.1\cdot 10^{-4}$, $x_{\rm HeII}=8.1\cdot 10^{-10}$ (an emergence of hydrogen molecules ${\rm H_2}$, ${\rm H_2^{+}}$  and negative hydrogen ions, ${\rm H^{-}}$, does not change these values essentially since
their number densities are of several orders lower than free electrons ones \citep{seager2000}). Residual values of ions fractions decrease with increasing of total 
baryon content. In fact, if $\Omega_b=0.06$ then at $z=200$  $x_{\rm e}\approx  x_{\rm HII}=6.3\cdot 10^{-4}$, $x_{\rm HeII}=2.0\cdot 10^{-10}$. Stronger dependence of residual values of helium ionized fraction  (HeII) on baryon content when compared to the  hydrogen one (HII) is explained by different 
 atomic structures of  HI and HeI atoms and conditions under which these recombinations occur: helium due to its higher potential of ionization starts
 to recombine earlier when number densities of plasma particles are higher and its recombination proceed in the conditions of high number
 density of free electrons caused by the complete hydrogen ionization.

\section{Perturbations of number densities of ions and electrons }

\subsection{Definitions}

Let the values of ionization fractions of HI, HII, HeI, HeII, HeIII and electrons averaged over the whole space at fixed cosmological time be  $x_i$, where {\it ''i''} marks each component among them. Let us denote as $\hat x_i$ the local value of relative number density of each component in the range of cosmological density perturbation of baryonic matter $\delta_{b}\equiv \delta \rho_b/ \rho_b\ll1$, were $\rho_b$ is its  matter density. Its deviation from mean 
value we mark as $\delta x_i$, so that $\hat x_i= x_i + \delta x_i$ and $\delta x_i$ is called the perturbation of relative number density of {\it i}-th component. Relative perturbations of relative number densities of ions and free electrons we define as 
$\Delta_i\equiv \delta x_i/ x_i$. It is obvious that
$\Delta_{\rm e}=\delta n_{e}/ n_{e} - \delta n_{\rm H}/ n_{\rm H}$, 
$\Delta_{\rm HII}=\delta n_{\rm HII}/ n_{\rm HII} - \delta n_{\rm H}/ n_{\rm H}$, 
$\Delta_{\rm HeII}=\delta n_{\rm HeII}/ n_{\rm HeII} - \delta n_{\rm He}/ n_{\rm He}$, 
$\Delta_{\rm HeIII}=\delta n_{\rm HeIII}/ n_{\rm HeIII} - \delta n_{\rm He}/ n_{\rm He}$.
We suppose that primordial chemical composition of  baryon matter is uniform ($f_{\rm He}$ is constant), so 
$\delta n_{\rm H}/ n_{\rm H}=\delta n_{\rm He}/ n_{\rm He}=\delta_{b}$ and 
\begin{equation}
\label{deltax}
\Delta_i=\delta_i -\delta_{b}, 
\end{equation}
where $\delta_i\equiv\delta n_{i}/ n_{i}=\Delta_i+\delta_{b}$ is relative number density perturbation of {\it i}-th component. It must be noted,
that in expanding universe the recombination does not end with the completely neutral hydrogen or helium, but with residual ionization
(see last paragraph of previous section). Therefore, none of values $n_i$ does reach zero and ambiguity of $"0/0"$-type  in $\delta_i$ does not appear.
So, $\Delta_i$'s do not take diverging values ever as it seen from (\ref{deltax}). Numerical results presented below prove that.

Therefore, $\Delta_i$ is 
difference of two relative perturbations: of number density of {\it ''i''}-th component and  density of all baryonic matter. Since $\delta_i$ and  $\delta_{b}$ are scalar functions of four coordinates in some gauge, so under the gauge transformations which do not change the cosmological background characteristics (mean CMB temperature, isotropic Hubble expansion etc.) each from them is transformed by  adding the same 
expression from component of transformation of time coordinate (see for details \citet{bardeen80,kodama84,durrer2001}). Whereas  in (\ref{deltax}) they appear with opposite signs then $\Delta_i$'s keep unchanged under such transformations, so they are gauge-invariant quantities. 

If ionization degree does not change with time, that is valid when hydrogen or helium are entirely ionized, then 
$\delta_i=\delta_{b}$ and $\Delta_i=0$. If the photorecombination and photoionization rates as well as ionization degree of some
component change, then  $\delta_i$ and $\delta_{b}$ can vary with different rates because the altering of $\delta_{b}$ is driven by 
gravitation and stress of baryon-photon plasma, while $\delta_i$ is additionally influenced by kinetics of ionization-recombination processes. Therefore,
$\Delta_i$ is measure of deviation of relative number density perturbations of {\it ''i''}-th component from relative density perturbation of total baryon component as a result of changing of recombination and ionization rates within cosmological density perturbation.

At enough early stage of evolution of the Universe for adiabatic density perturbations of scales larger than the horizon $\delta_{m}=\delta_{b}$ and relative perturbations of radiation energy density $\delta_{\rm R}\equiv \delta \epsilon_{\rm R}/ \epsilon_{\rm R}=4\delta_{b}/3$. Since  $\epsilon_{\rm R}=aT_{\rm R}^4$, then $\delta_{T_R}\equiv \delta T_R/ T_R=1/3\delta_b$. For isothermal perturbations $\delta_{T_R}=0$. 

\subsection{Equations}

Long before and much after recombination the local baryon mass density fluctuations most probably  lead  to corresponding perturbations of  number densities of ions and electrons, $ \delta_i\approx \delta_b$. But at the recombination epoch because of dependence of ionization-recombination process rates on density and temperature  of baryon matter  the distribution of atoms over ionization stages within those perturbations will  somewhat 
depart from background one and  $\Delta_i\ne 0$  is expected to be true.  We study the cosmological perturbations of small amplitudes, so the ratio
$t_{\rm Hubble}:t_{\rm T}:t_{\rm HI}:t_{\rm HeI}$ remains practically the same as for background. It means that within cosmological perturbations the same equations (\ref{SahaHeIII})-(\ref{Tm2}) are applicable and connection between the perturbations of ion number density  and cosmological perturbations of density and temperature may be obtained by  variation of those equations.

Varying the variables $n_{\rm H}$, $T_m$, $x_{\rm HeII}$, $x_{\rm HeIII}$ and $x_e$  in the equation (\ref{SahaHeIII}) we will obtain:
\begin{eqnarray}
\label{SahadxHeIII}
\Delta_{\rm HeIII} = {\displaystyle x_{\rm e}(1-x_{\rm HeIII})\over\displaystyle  x_{\rm e}+(1-x_{\rm HeIII})x_{\rm HeIII}f_{\rm He}}
\left[\left({\displaystyle 3\over\displaystyle  2} +{\displaystyle \chi_{\rm HeII}\over \displaystyle kT_m}\right)\delta_{T_m} -\delta_{b}\right],\;\;\;\; 
\Delta_{\rm HeII}=-\Delta_{\rm HeIII}, \;\;\; \Delta_{\rm e}={\displaystyle x_{\rm HeIII}\over\displaystyle  x_{\rm e}}f_{\rm He}\Delta_{\rm HeIII},
\end{eqnarray}
Here it is assumed that $x_{\rm HI}=x_{\rm HeI}=0$ at $z>3500$  and relative perturbations of rest of components have vanished too. One can see that
relative perturbations of relative number densities of ions HeIII is the linear combination of cosmological perturbations of temperature and mass density of baryon matter. Within an adiabatic perturbation $\Delta_{\rm HeIII}$ has the same sign as temperature fluctuation and opposite to mass density one.  The values of  $x_{\rm e}$ and $x_{\rm HeIII}$ are calculated from (\ref{SahaHeIII}). The asymptotical behaviour of 
$\Delta_{\rm HeIII}$ follows from (\ref{SahadxHeIII}): at $z>7000$ when $x_{\rm HeIII}\to 1$ (all helium atoms become  double ionized)  $\Delta_{\rm HeIII}\to 0$ ($\delta_{\rm HeIII}=\delta_{b}$). On the other hand, at $z<5000$ when $x_{\rm HeIII}\to 0$  $\Delta_{\rm HeIII}\to  {\chi_{\rm HeII}\over kT_m}\delta_{T_m}$ and increases with temperature decreasing. It is obvious that the second asymptotics is not physically correct. Indeed, such  monotonous increasing of $\Delta_{\rm HeIII}$ is caused by vanishing of $x_{\rm HeIII}$ (see Fig.\ref{rec-Tm}) and does not describe the real number density perturbation of  HeIII ions. That is why in Fig.\ref{pos-neg} in the range of $5000<z<7000$ the absolute values of perturbations of number density $\delta{x_{\rm HeIII}}$  and  $\delta{x_{\rm HeII}}$ are presented. 

At $3500<z<5000$ both hydrogen and helium are entirely ionized: $x_{\rm HII}=x_{\rm HeII}=1$, $x_{\rm HI}=x_{\rm HeI}=x_{\rm HeIII}=0$. So, at this period the amplitudes of all relative perturbations equal to zero. With subsequent decreasing of temperature 
the HeI atoms and afterwards HI ones begin to recombine. The kinetics of their recombination is described by Saha equations  (\ref{SahaHeII}) and (\ref{SahaHII}). Variation of these equations gives the following expressions for relative perturbations of relative number density of helium 
$\Delta_{\rm HeII}$, hydrogen $\Delta_{\rm HII}$ and free electrons $\Delta_{\rm e}$: 
\begin{eqnarray}
\label{SahadxHeII}
\Delta_{\rm HeII} = (1-x_{\rm HeII}){(1-x_{\rm HII})x_{\rm HII}\left({\chi_{\rm HI}\over kT_m}-{\chi_{\rm HeI}
\over kT_m}\right)\delta_{\rm T_m}+x_{\rm e}\left[\left({3\over 2}+{\chi_{\rm HeI}\over kT_m}\right)
\delta_{\rm T_m}-\delta_b\right]\over (1-x_{\rm HII})x_{\rm HII}+(1-x_{\rm HeII})x_{\rm HeII}f_{\rm He}-x_{\rm e}},
\end{eqnarray}
\begin{eqnarray}
\label{SahadxHII}
\Delta_{\rm HII} = {(1-x_{\rm HII})x_{\rm e}\left[\left({3\over 2}+{\chi_{\rm HI}\over kT_m}
\right)\delta_{\rm T_m}-\delta_b\right] \over (1-x_{\rm HII})x_{\rm HII}+(1-x_{\rm HeII})x_{\rm HeII}f_{\rm He}-x_{\rm e}},
\end{eqnarray}
\begin{eqnarray}
\label{Sahadxe}
\Delta_{\rm e} = {(1-x_{\rm HII})x_{\rm HII}{\chi_{\rm HI}\over kT_m}\left(1+(1-x_{\rm HeII})x_{\rm HeII}/x_{\rm e}\right)\over
(1-x_{\rm HII})x_{\rm HII}+(1-x_{\rm HeII})x_{\rm HeII}f_{\rm He}-x_{\rm e}}\delta_{\rm T_m}
+{(1-x_{\rm HeII})x_{\rm HeII}{\chi_{\rm HeI}\over kT_m}\left(f_{\rm He}-(1-x_{\rm HII})x_{\rm HII}/x_{\rm e}\right)\over
(1-x_{\rm HII})x_{\rm HII}+(1-x_{\rm HeII})x_{\rm HeII}f_{\rm He}-x_{\rm e}}\delta_{\rm T_m}+  \nonumber \\
+ {(1-x_{\rm HII})x_{\rm HII}+(1-x_{\rm HeII})x_{\rm HeII}f_{\rm He}\over 
(1-x_{\rm HII})x_{\rm HII}+(1-x_{\rm HeII})x_{\rm HeII}f_{\rm He}-x_{\rm e}}\left[{3\over 2} \delta_{\rm T_m}-\delta_b\right]
\end{eqnarray}
Their asymptotic behaviour for $x_{\rm HeII}\to 1$ and  $x_{\rm HII}\to 1$ agrees with our anticipations and appears to be the same as for 
 $\Delta_{\rm HeIII}$ case: $\Delta_{\rm HeII}$,  $\Delta_{\rm HII}$ and $\Delta_{\rm e}$ $\to 0$. 
Other asymptotics for $x_{\rm HeII}\to 0$  and  $x_{\rm HII}\to 0$  have not physical sense, since 
for $x_{\rm HeII}\le 0.99$ and $x_{\rm HII}\le 0.99$ it is necessary to use the non-equilibrium rate equations and energy balance (\ref{eqHeII})-(\ref{Tm2}). In this case the differential equations for relative perturbations $\Delta_{\rm HII}$, $\Delta_{\rm HeII}$  and $\delta_{\rm T_m}$ are obtained by variation of (\ref{eqHeII})-(\ref{Tm2}):
\begin{eqnarray}
\label{eqdxHeII}
&&x_{\rm HeII}{\displaystyle d\Delta_{\rm HeII}\over\displaystyle  dz} ={\displaystyle \left(1 + K_{\rm HeI} \Lambda_{\rm He} n_{\rm H}
  (1-x_{\rm HeII}){\rm e}^{-h\nu_{ps}/kT_{\rm m}}\right)x_{\rm HeII}x_{\rm e} n_{\rm H} \alpha_{\rm HeII}
    \over\displaystyle  H(z)(1+z)\left(1+K_{\rm HeI}  (\Lambda_{\rm He} + \beta_{\rm HeI}) n_{\rm H} (1-x_{\rm HeII})
  {\rm e}^{-h\nu_{ps}/kT_{\rm m}}\right)}\left[\Delta_{\rm e}+\Delta_{\rm HeII}+
  \delta_b+{\delta \alpha_{\rm HeI}\over \alpha_{\rm HeI}}\right]- \nonumber \\ 
&&-{\displaystyle \left(1 + K_{\rm HeI} \Lambda_{\rm He} n_{\rm H} (1-x_{\rm HeII}){\rm e}^{-h\nu_{ps}/kT_{\rm m}}\right)\beta_{\rm HeI} 
  (1-x_{\rm HeII}) {\rm e}^{-h\nu_{\rm HeI2^1s}/kT_{\rm m}}\over \displaystyle H(z)(1+z)\left(1+K_{\rm HeI}
  (\Lambda_{\rm He} + \beta_{\rm HeI}) n_{\rm H} (1-x_{\rm HeII})
  {\rm e}^{-h\nu_{ps}/kT_{\rm m}}\right)} \left[{\delta \beta_{\rm HeI}\over \beta_{\rm HeI}}-
  {x_{\rm HeII}\over 1-x_{\rm HeII}}\Delta_{\rm HeII}+{h\nu_{\rm HeI2^1s}\over kT_{\rm m}}\delta_{T_{\rm m}}\right]+\nonumber \\
&&+{\displaystyle dx_{\rm HeII}\over\displaystyle  dz}{\displaystyle K_{\rm HeI}\Lambda_{\rm He} n_{\rm H} (1-x_{\rm HeII})
  {\rm e}^{-h\nu_{ps}/kT_{\rm m}}\over\displaystyle  1+K_{\rm HeI}\Lambda_{\rm He}n_{\rm H} (1-x_{\rm HeII})
  {\rm e}^{-h\nu_{ps}/kT_{\rm m}}}\left[\delta_b-{x_{\rm HеII}\over 1- x_{\rm HеII}}\Delta_{\rm HeII}+
  {h\nu_{ps}\over kT_{\rm m}}\delta_{T_{\rm m}}\right]-\Delta_{\rm HeII}{\displaystyle dx_{\rm HeII}\over\displaystyle  dz}-  \\ 
&&-{\displaystyle dx_{\rm HeII}\over \displaystyle dz}{\displaystyle K_{\rm HeI}  (\Lambda_{\rm He} + \beta_{\rm HeI}) n_{\rm H} (1-x_{\rm HeII})
  {\rm e}^{-h\nu_{ps}/kT_{\rm m}}\over\displaystyle  1+K_{\rm HeI}  (\Lambda_{\rm He} + \beta_{\rm HeI}) n_{\rm H} (1-x_{\rm HeII})
  {\rm e}^{-h\nu_{ps}/kT_{\rm m}}}\left[\delta_b-{x_{\rm HII}\over 1- x_{\rm HII}}\Delta_{\rm HeII}+
  {h\nu_{ps}\over kT_{\rm m}}\delta_{T_{\rm m}}+{\beta_{\rm HeI}\over \Lambda_{\rm He} + \beta_{\rm HeI}}
  {\delta \beta_{\rm HeI}\over \beta_{\rm HeI}}\right],  \nonumber
 \end{eqnarray}
 
 \begin{eqnarray}{}
\label{eqdxHII}
&&x_{\rm HII}{\displaystyle d\Delta_{\rm HII}\over\displaystyle  dz} ={\displaystyle \left(1 + K_{\rm HI} \Lambda_{\rm H} n_{\rm H}
  (1-x_{\rm HII})\right)x_{\rm HII}x_{\rm e} n_{\rm H} \alpha_{\rm HII}
    \over\displaystyle  H(z)(1+z)\left(1+K_{\rm HI}  (\Lambda_{\rm H} + \beta_{\rm HI}) n_{\rm H} (1-x_{\rm HII})
  \right)}\left[\Delta_{\rm e}+\Delta_{\rm HII}+\delta_b+{\delta \alpha_{\rm HI}\over \alpha_{\rm HI}}\right]+\nonumber \\
&&+{\displaystyle \left(1 + K_{\rm HI} \Lambda_{\rm H} n_{\rm H} (1-x_{\rm HII})\right)\beta_{\rm HI} 
  (1-x_{\rm HII}) {\rm e}^{-h\nu_{\rm H 2s}/kT_{\rm m}}\over\displaystyle  H(z)(1+z)\left(1+K_{\rm HI}
  (\Lambda_{\rm H} + \beta_{\rm HI}) n_{\rm H} (1-x_{\rm HII})\right)} \left[{\delta \beta_{\rm HI}\over \beta_{\rm HI}}-
  {x_{\rm HII}\over 1-x_{\rm HII}}\Delta_{\rm HII}+{h\nu_{\rm HI2s}\over kT_{\rm m}}\delta_{T_{\rm m}}\right]+\nonumber \\
&&+{\displaystyle dx_{\rm HII}\over\displaystyle  dz}{\displaystyle K_{\rm HI}\Lambda_{\rm H} n_{\rm H} (1-x_{\rm HII})
 \over\displaystyle  1+K_{\rm HI}\Lambda_{\rm H} n_{\rm H} (1-x_{\rm HII})}\left[\delta_b-{x_{\rm HеII}\over 1- x_{\rm HеII}}\Delta_{\rm 
 HII}\right]-\Delta_{\rm HII}{dx_{\rm HII}\over dz}-\\
 &&\displaystyle -{\displaystyle dx_{\rm HII}\over\displaystyle  dz}{\displaystyle K_{\rm HI}  (\Lambda_{\rm H} + \beta_{\rm HI}) n_{\rm H} 
 (1-x_{\rm HII})\over\displaystyle  1+K_{\rm HI}  (\Lambda_{\rm H} + \beta_{\rm HI}) n_{\rm H} (1-x_{\rm HII})
}\left[\delta_b-{x_{\rm HII}\over 1- x_{\rm HII}}\Delta_{\rm HII}+{\beta_{\rm HI}\over \Lambda_{\rm H} + \beta_{\rm HI}}
  {\delta \beta_{\rm HI}\over \beta_{\rm HI}}\right], \nonumber
 \end{eqnarray}
 \begin{eqnarray}
\label{eqdTm}
 &&T_{\rm m}\frac{\displaystyle d\delta_{T_{\rm m}}}{\displaystyle dz} = \frac{\displaystyle 8\sigma_{\rm T}a_{\rm R}
   T_{\rm R}^4}{\displaystyle 3H(z)(1+z)m_{\rm e}c}\,
  \frac{\displaystyle x_{\rm e}}{\displaystyle 1+f_{\rm He}+x_{\rm e}} 
 \left[4(T_{\rm m} - T_{\rm R})\delta_{T_{\rm m}}+
  \frac{\displaystyle 1+f_{\rm He}}{\displaystyle 1+f_{\rm He}+x_{\rm e}}(T_{\rm m} - T_{\rm R})\Delta_{\rm e}
  +T_{\rm m}\delta_{T_{\rm m}}- T_{\rm R}\delta_{T_{\rm R}}\right]+ \nonumber \\
  &&+\frac{\displaystyle 2T_{\rm m}}{\displaystyle 1+z}\delta_{T_{\rm m}}-\frac{\displaystyle dT_{\rm m}}{\displaystyle dz}\delta_{T_{\rm m}},
  \end{eqnarray}
where ${\displaystyle dx_{\rm HeII}\over\displaystyle  dz}$, ${\displaystyle dx_{\rm HII}\over\displaystyle  dz}$ and  
$\frac{\displaystyle dT_{\rm m}}{\displaystyle dz}$ mean the same as the right-hand side of 
(\ref{eqHeII}),  (\ref{eqHII}) and  (\ref{Tm2}) respectively. The variations of recombination coefficients are expressed via temperature 
perturbations by the equations:
$${\delta \alpha_{\rm HI}\over \alpha_{\rm HI}}=\left(b-{d\cdot c\cdot t^d\over 1+c\cdot t^d}\right)\delta_{T_{\rm m}},$$ 
$${\delta \alpha_{\rm HeI}\over \alpha_{\rm HeI}}= -{1\over 2}\left(1+{(1-p)\sqrt{T_{\rm m}/T_2}\over 1+\sqrt{T_{\rm m}/T_2}}+
{(1+p)\sqrt{T_{\rm m}/T_1}\over 1+\sqrt{T_{\rm m}/T_1}}\right)\delta_{T_{\rm m}},$$
Then the variations of photoionization rates can be calculated in the following way:
$${\delta \beta_i \over \beta_i}= {\delta \alpha_i \over \alpha_i }+{3\over 2}\delta_{T_{\rm m}}+{h\nu_{i 2s}\over kT_{\rm m}}\delta_{T_{\rm m}}.$$ 

So, equations (\ref{eqdxHeII})-(\ref{eqdTm}) constitute the system of three ordinary linear differential equations of first order  for relative perturbations of ions and electrons relative number densities and matter temperature, which can be solved using the publicly available code DVERK\footnote[1]{It is created by T.E. Hull, W.H.Enright, K.R. Jakson  in 1976 and is available at site http://www.cs.toronto.edu/NA/dverk.f.gz}. The initial data for them are equilibrium values of relative perturbations of  ion relative number density at the moment when $x_{\rm HII}>0.99$  and $x_{\rm HeII}>0.99$ 
calculated by (\ref{SahadxHeII})-(\ref{SahadxHII}).

Let us use the equations (\ref{SahadxHeIII})-(\ref{eqdTm}) to analyse the evolution of relative ion density perturbations and temperature of baryonic matter. Since all these equations have solutions in unperturbed problem therefore it seems naturally to supplement the code RECFAST \citep{seager1999} with block for calculation of perturbations of ionized fractions. The complemented code $drecfast.f$
\footnote[2]{available at http://astro.franko.lviv.ua/$\sim$novos/} 
is used further in our analysis of perturbations of ion number density and matter temperature.

\subsection{Results}

To estimate the  magnitude of probable effect it is useful  first to consider  the stationary adiabatic mass density perturbation of baryon matter  with
some amplitude. Its spatial shape does not matter for our analysis.  One may assume  for simplicity that it is homogeneous in some region of space.

\begin{figure}
\centerline{\includegraphics[width=8cm]{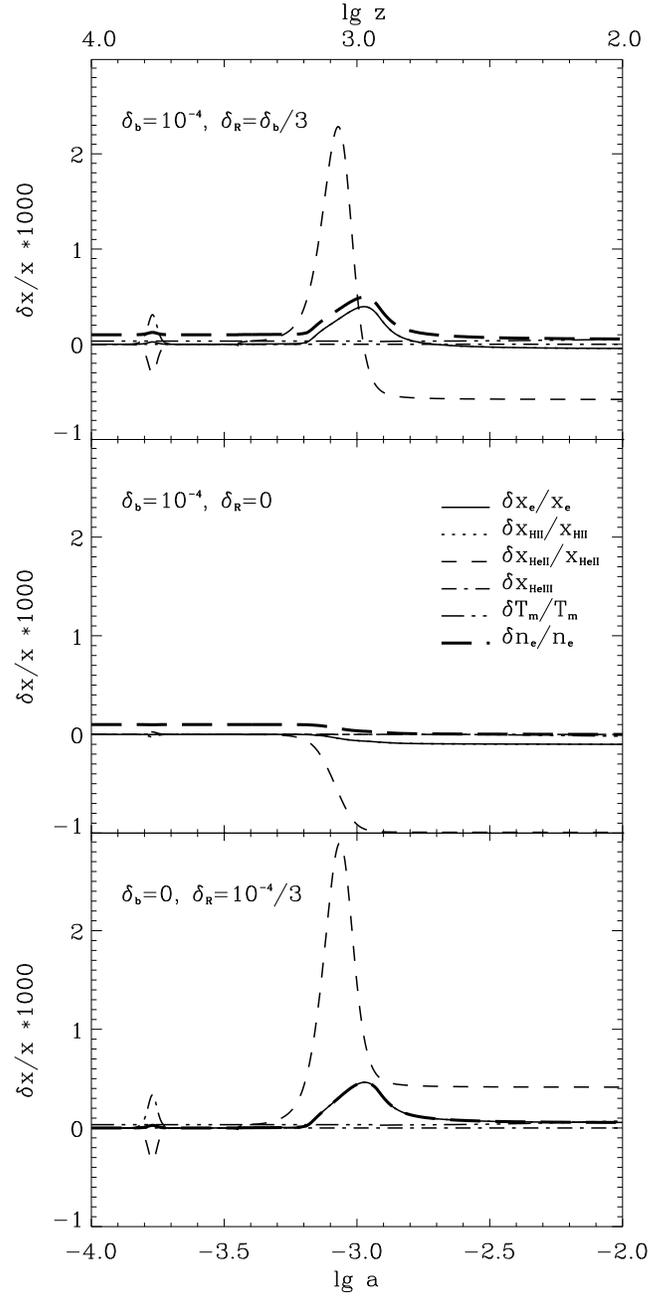}}
\caption{The relative number density perturbations of ions of helium, hydrogen and free electrons caused by adiabatic positive 
(overdensity) matter density initial perturbation (top panel),  by isothermal  perturbation (middle panel) and by hot ''thermal'' initial  fluctuation (bottom panel).}
\label{pos-neg}
\end{figure}

The results of calculations of ion number density relative perturbations  (\ref{SahadxHeIII})-(\ref{eqdTm}) caused by adiabatic positive 
matter density perturbation (overdensity) with $\delta_b=10^{-4}$ and $\delta_{T_{\rm R}}=\delta_b/3$ as well as isothermal  ($\delta_b=10^{-4}$, $\delta_{T_{\rm R}}=0$) and ''thermal''  ($\delta_b=0$, $\delta_{T_{\rm R}}={1\over 3}\cdot 10^{-4}$) ones are presented  in Fig.\ref{pos-neg} for the range of redshifts $200<z<10000$. One can see  that the  appreciable deviations of relative perturbations of ion number density $\Delta_{\rm HII}$, $\Delta_{\rm HeII}$ and $\Delta_{\rm e}$ from perturbation of total baryon mass density $\delta_b$  arise in cosmological recombination epoch ($800<z<1500$) as result of  amplification of photorecombination and photoionization rates. So, the amplitudes of relative perturbations of number densities of protons and electrons  are of $\simeq 5$ times higher than the amplitude of total baryon number density perturbation. For helium ions HeII such ratio is even higher, $\simeq 18$. However,  such large amplitude practically does not contribute to the number density perturbations of free electrons: in the Fig.\ref{pos-neg} $\Delta_{\rm HII}$ and $\Delta_{\rm e}$ are superposed in the range of $200<z<1500$. This is because of  too low number density of helium ions. Indeed, at  $z\simeq 1200$, when $x_{\rm HII}\simeq 0.5$, $x_{\rm HeII}<0.0003$. Also one can see, that relative temperature perturbations are equal to the initial ones because of dominant CMB radiation in the energy balance of recombinational plasma.  So, the fluctuations in ionization-recombination rates caused by cosmological adiabatic perturbations do not  result into additional noticeable local fluctuations of CMB and electron-ion temperatures.  

The figures in the middle and bottom panels illustrate the role of the initial mass density perturbation and initial temperature perturbation in arising 
of perturbations of number densities of ions and electrons. 

For the adiabatic negative (underdensity) perturbations as well as for the isothermal  underdensity and cold ''thermal'' ones the pictures are symmetric as follows from equations   (\ref{SahadxHeIII})-(\ref{eqdTm}).

\section{Perturbations of number density of electrons in $\Lambda$CDM model}

In previous section we have analysed the perturbations of number density of ions in the framework of ''toy'' model of stationary perturbations. In reality,  
the amplitudes of adiabatic cosmological perturbations evolve due to gravitational attraction and repulsion by stress in baryon-photon plasma in potential wells created by the density perturbations of dark matter, see for example \citet{bardeen80,kodama84,ma1995,hu1995,durrer2001} and citing therein.  When the scale of perturbation becomes substantially smaller than scale of acoustic horizon (Jeans scale), then  adiabatic perturbations in the baryon-photon plasma start to oscillate like the standing acoustic waves. In consequence of recombination the Jeans scale drops and the previously oscillating amplitudes of perturbations in baryon component start to monotonously increase mainly under influence of gravitational attraction of dark matter density perturbations. The 
amplitudes of perturbations with scale larger than acoustic horizon at recombination epoch increased  $\delta_b\propto t^{1/2}$ in radiation-dominated epoch and $\delta_b\propto t^{2/3}$ after recombination in dust-like Universe. In  papers  \citep{bardeen80,kodama84,ma1995,hu1995,durrer2001} one can find analytical solutions of relevant equations for evolution of relative
 density perturbations in simplified cases of single component mediums as well as the numerical solutions for realistic multi-component Universe. 
 
\begin{figure}
\centerline{\includegraphics[width=8cm]{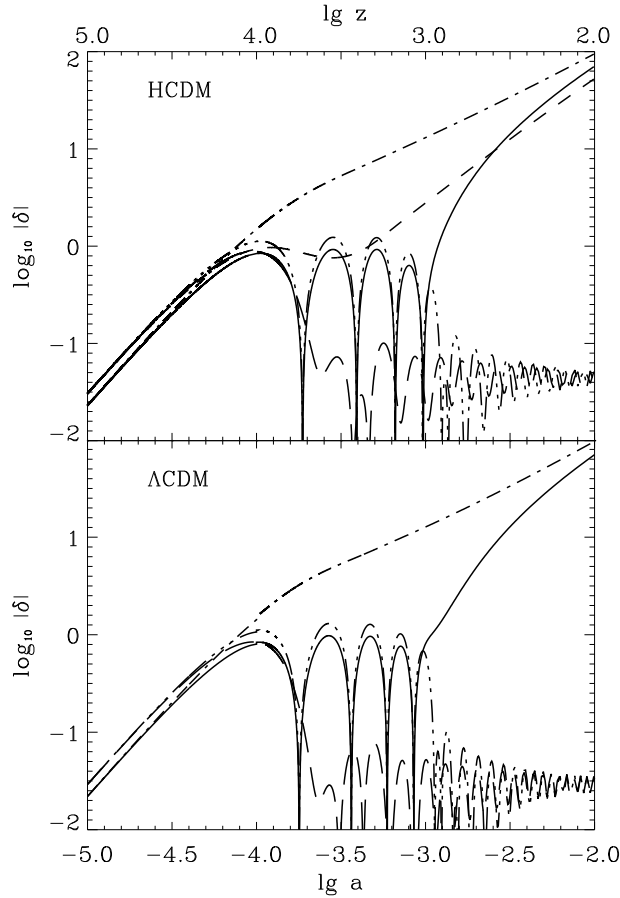}}
\caption{Evolution of amplitudes of adiabatic density perturbations for different components of  HCDM 
 ($\Omega_b=0.05$, $\Omega_{\rm CDM}=0.75$, $\Omega_{\nu}=0.2$, $h=0.65$) and $\Lambda{\rm CDM}$ 
  ($\Omega_b=0.05$, $\Omega_{\rm CDM}=0.3$, $\Omega_{\Lambda}=0.65$, $h=0.65$)  multi-component models in synchronous gauge
 as calculated by COSMICS code (arbitrary normalization). The lines represent the relative density perturbation of baryons $\delta_b$ (solid line), thermal 
  electromagnetic radiation $\delta_{\rm R}$ (dash-three dotted line), cold dark matter $\delta_c$ (dash-dotted line), massless (long-dashed line) 
  and massive (short-dashed line) neutrino. The scale of  perturbations  in Fourier space is $k=0.1$Mpc$^{-1}$.} 
\label{deltanma}
\end{figure}

 I shall use here the numerical approach by \citet{ma1995} and their package of FORTRAN programs COSMICS\footnote[3]{It is available at site 
 http://arcturus.mit.edu/cosmics/}  in order to calculate the amplitude of baryon density perturbations in 
 multi-component medium for synchronous gauge. The evolution of amplitudes of adiabatic density perturbations  for each component of  
 Hot plus Cold  Dark Matter (HCDM) and $\Lambda{\rm CDM}$  models are shown in the Fig.\ref{deltanma}. The scale of these perturbations 
 in Fourier space is $k=0.1$Mpc$^{-1}$, that actually equals to the horizon distance at the epoch of  radiation-matter equality 
 and amounts approximately 1/4 of the horizon size at the decoupling epoch. So, starting from radiation-matter equality epoch 
 ($a_{eq}\simeq 2\cdot 10^{-4}\Omega_mh^2$) the amplitudes of density perturbations in baryon-photon plasma oscillate acoustically till the 
 recombination ($a_{rec}\simeq  10^{-3}$). The detailed analysis of the evolution of density perturbations of different scales in the
  multi-component Universe one can find in  \citet{ma1995}.  

\begin{figure*}
\centerline{\includegraphics[width=18cm]{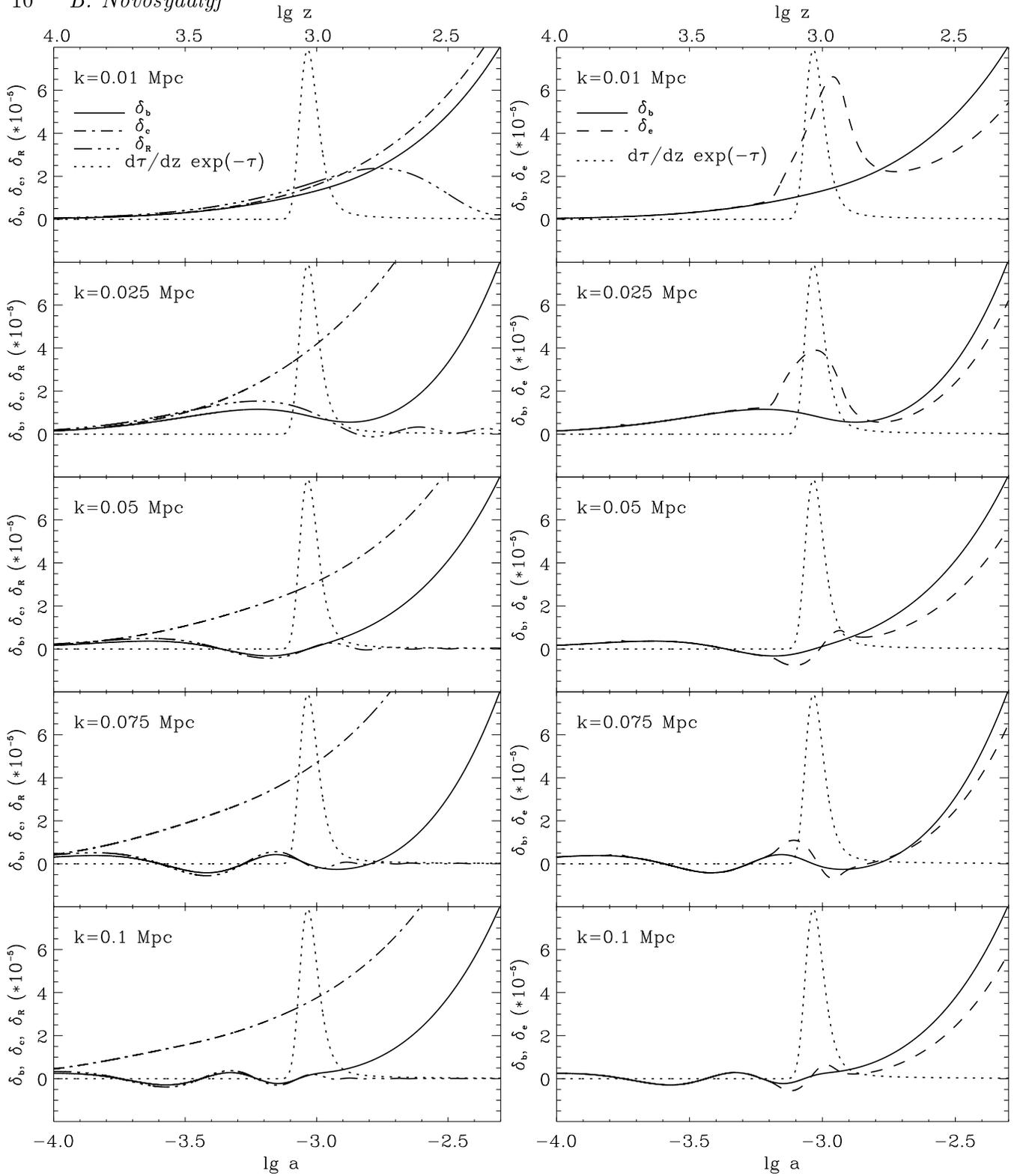}}
\caption{Left-hand column: evolution of relative density perturbations $\delta_b$, $\delta_{\rm R}$ and $\delta_c$ for the baryon 
(solid line), photon (dash-three dotted line) and CDM (dash-dotted line) components  with adiabatic initial conditions and free normalization.  
Right-hand column: electron number density perturbations against baryon mass density ones. The amplitudes of all perturbations are in dimensionless units $10^{-5}$. The visibility function $d\tau /dz e^{-\tau}$ (multiplied by 0.018 for convenience, dotted line) denotes the position of last scattering surface.} 
\label{deltan}
\end{figure*}

In order to analyse the perturbations of number densities of ions and electrons caused by cosmological baryon density perturbations the code {\it drecfast.f} was complemented by the COSMICS' code {\it linger\_syn.f} as subroutine, so, that amplitude of baryon  mass density perturbation $\delta_b$  and radiation temperature $\delta_{T_{\rm R}}=\delta_{\rm R}/4$ were precalculated at each step of integration of the equations system  (\ref{SahadxHeIII})-(\ref{eqdTm}). Results of joint calculations the  ion number density  and mass density perturbations of different scales in  $\Lambda{\rm CDM}$ model are presented in Fig.\ref{deltan}. In the left-hand column the evolution of relative mass density perturbations $\delta_b$, $\delta_{\rm R}$ and $\delta_c$ for the baryon, photon and CDM components  (adiabatic initial conditions) are shown. Five wave 
numbers are plotted:  $k=0.01,\;0.025,\;0.05,\;0.075,\;0.1$ Mpc$^{-1}$ (from top to bottom). In the right-hand column the perturbations of  number density of electrons against baryon mass density ones are shown for the same wave numbers. In each figure the visibility function 
$d\tau /dz e^{-\tau}$ is shown by dotted line in order to mark the position of last scattering surface. The peak of this function is found at $z=1088$ and, in fact, denotes moment of decoupling  of photons and baryons. The particle horizon at this moment equals $\eta_{dec}\simeq 278$Mpc
($k_{dec}\simeq 0.023$Mpc$^{-1}$), sound horizon (or Jeans scale) is  $\lambda^s_{dec}\simeq 160$Mpc  ($k^s_{dec}\simeq 0.039$Mpc$^{-1}$). Therefore, the wave numbers of particle and sound horizons at decoupling epoch fall within the range of plotted $k$ and 
behaviour of baryon mass density and free electron number density perturbations of scales larger, comparable and lower in comparison with horizons
 are revealed. 
For modes with $k<0.01$Mpc$^{-1}$ the relative amplitude, shape and position of  peak of electron number density perturbation are the same as 
for $k=0.01$Mpc$^{-1}$ mode. For them the peak of $\delta_{\rm e}$  is situated at $900\le z\le 910$ and ratio of amplitudes 
$\delta_{\rm e}/\delta_b\simeq 4.5$. For modes with $k>0.01$Mpc$^{-1}$ the relative amplitude, shape and position of  peak of electron number density perturbation vary: when a wave number increases starting from 0.01 to $k^s_{dec}$ the peak position $\delta_{\rm e}$ shifts to the position of visibility function peak with approximately the same ratio
of $\delta_{\rm e}/\delta_b\approx 4.5$. For modes with  $k>k^s_{dec}$ the value of  $\delta_{\rm e}$ radically decreases and makes several oscillation around 
value of  $\delta_b$ depending on phase of $\delta_{\rm R}$ oscillation at decoupling epoch.  
So, ion number density fluctuations at decoupling epoch caused by recombination kinetics in the range of adiabatic density perturbations are most 
prominent for the  large-scale perturbations with  $k\le k^s_{dec}$.
At $z<800$, when ionization become residual, $\delta_{\rm e}$  becomes lesser than $\delta_b$  independently of scale of baryon density perturbation.  
The difference  acquires the value $\sim 20\%$ at $z\sim 200$.  It is caused by increasing of $\delta_b$ on all scales under gravitational potential of dark matter density perturbations, so, the probability of  recombination is slightly higher in such regions. It is similar to the residual ionization and its  inverse dependence on baryon density. 

Finally it should be noted that presented results do not depend essentially on exact values of cosmological parameters of non-exotic models and are 
practically the same for parameters of  WMAP concordance model. 

\section{$\Delta_{\rm e}$ and corrugation of last scattering surface}

The results presented in Fig.\ref{pos-neg} prove that ion number density fluctuations caused by kinetics of ionization-recombination processes in the field of cosmological adiabatic perturbations do not lead to appreciable additional local temperature fluctuations. This was expected  
because of  dominating role of the thermal (relic) electromagnetic radiation in the energy balance of photon-baryon plasma.  However, to calculate
 of  CMB temperature fluctuations and polarization at high accuracy in order to achieve better agreement between theory and 
observations it is necessary to take into account  the contributions of such electron number density fluctuations into optical depth  $\tau$ caused 
by Thomson scattering . 

The visibility function $d\tau /dz e^{-\tau}$, where  $\tau(z)=\int_0^z {c\sigma_{\rm T} n_{\rm e}(z) H^{-1}(z)(z+1)^{-1}dz}$, represents the probability that a photon was scattered for the last time within $dz$ of $z$. So, the main part of CMB photons come to observer from the vicinity of  maximum of visibility function. The fluctuations of  number density of electrons result into faint ''corrugation'' of last scattering surface of thermal relic radiation:  the maximum of visibility function  will be at somewhat lower redshifts for overdensity perturbations  and at somewhat higher redshifts for underdensities
 than for  unperturbed region.  We can estimate this effect in the following way. 

The redshift of peak can be determined from the condition of local extremum giving the equation:
\begin{eqnarray}
\label{zdec}
{1\over x_e}{d x_e\over dz}-{1\over H(z)}{d H(z)\over dz} + {2\over z+1}- {c\sigma_{\rm T}n_{\rm H}x_e(z) \over H(z)(z+1)}=0,
\end{eqnarray}
where $n_e(z)$ is unperturbed  number density of electrons shown in Fig.\ref{rec-Tm}, 
$H(z)=H_0\left[\Omega_{\rm R}(z+1)^4+\Omega_m(z+1)^3+\Omega_k(z+1)^2+\Omega_{\Lambda}\right]^{1/2}$  is Hubble constant.
Its solution for $z$ gives a position of peak of visibility function in unperturbed medium. For $\Lambda$CDM model with the same parameters as
for Fig.\ref{deltan} using the numerical solution we obtain $z_{dec}\simeq 1088$. In the range of perturbation, where 
$\hat n_{\rm H}=n_{\rm H}(1+\delta_b)$ and $\hat x_e=x_e(1+\Delta_{\rm e})$, it is expected to be $\hat z_{dec}=z_{dec}+\delta z_{dec}$. 
The displacement of peak $\delta z_{dec}$ can be estimated by variation of equation (\ref{zdec}) and expanding of functions $x_e({\hat z_{dec}})$ and $n_{\rm H}({\hat z_{dec}})$ into Taylor series about $z_{dec}$ (in linear approximation):
\begin{eqnarray}
\label{dzdec}
&&{\displaystyle \delta z\over\displaystyle   z+1}=\left[{\displaystyle c\sigma_{\rm T}n_{\rm H}x_e(z) \over\displaystyle  H(z)(z+1)}(\delta_b+\Delta_{\rm e})-{\displaystyle d\Delta_{\rm e}\over\displaystyle  dz}\right]\nonumber \\
&&\times\left[{c\sigma_{\rm T}n_{\rm H}x_e(z) \over\displaystyle  H(z)(z+1)}\left({\displaystyle 1\over\displaystyle  H(z)}{\displaystyle d H(z)\over\displaystyle dz}(z+1)-{\displaystyle 1\over\displaystyle  x_e}{\displaystyle d x_e\over\displaystyle dz}(z+1)-2\right)-(z+1){\displaystyle d\over \displaystyle dz}\left({\displaystyle 1\over\displaystyle  H(z)}{\displaystyle d H(z)\over\displaystyle  dz}-{\displaystyle 1\over\displaystyle  x_e}{\displaystyle d x_e\over \displaystyle dz}\right)-{\displaystyle 2\over\displaystyle  z+1}\right]^{-1}.
\end{eqnarray}
Therefore, such displacement depends on the amplitude of $\Delta_{\rm e}+\delta_b=\delta_e$ and the gradient of $\Delta_{\rm e}$. For arbitrary normalisation of amplitude of cosmological perturbations it is conveniently to present the result in units of relative density perturbations of  baryon-photon plasma components and for large-scale perturbations ($k<k^s_{dec}$) we have:  $\delta z_{dec}/(z_{dec}+1)\approx -0.25\delta_{\rm R}=-0.33\delta_b$. If we suppose $\Delta_{\rm e}=0$ (ion number density perturbations follow the baryon mass density ones) then $\delta z_{dec}/(z_{dec}+1)\approx -0.051\delta_{\rm R}=-0.068\delta_b$. So, the amplification of perturbation amplitudes of electron number density  by recombination processes makes the last scattering surface more ''corrugated'' in optical depth. Such ''corrugation'' results into observable CMB temperature fluctuations.

We can make the estimation of this effect. Let us suppose that major part of CMB photons come from thin last scattering surface placed at $z_{dec}$ from unperturbed medium and $\hat z_{dec}=z_{dec}+\delta z_{dec}$  from perturbed one. Because  temperature of the thermal radiation at $z_{dec}$ and at 
$\hat z_{dec}$ is the same then its observable variation follows from the well-known relation $T_0=T_{dec}/(z_{dec}+1)$:
$\left(\delta T_0/T_0\right)_{cor}=-\delta z_{dec}/(z_{dec}+1)\approx 0.25\delta_{\rm R}=0.33\delta_b$ that equals to intrinsic adiabatic CMB temperature fluctuation, $\left(\delta T_0/T_0\right)_{ad}= \delta_{\rm R}/4=\delta_b/3$, exactly. It is other way for calculation of adiabatic term of CMB primary anisotropy, which, however, requires the accurate precalculation of $\Delta_e$ or $\delta_e$ different from $\delta_b$ at last scattering surface. When we suppose that $\delta_e=\delta_b$ ($\Delta_e=0$) then $\left(\delta T_0/T_0\right)_{cor}\approx 0.2 \left(\delta T_0/T_0\right)_{ad}$ that is incorrect.
 
 For small-scale perturbations $k>k^s_{dec}$, the value of $\delta z_{dec}/(z_{dec}+1)$ will be  smaller and its sign will 
  alternate depending on oscillation phases of $\delta_{\rm R}$ and $\delta_b$ at $z_{dec}$. The assumption of thin LSS  is too rough in this case to 
  make correct estimation. The line-of-sight integration must be undertaken here because of fuzziness effect leading to exponential reduction of 
  any primary CMB  anisotropy. Unfortunately, straightforward substitution of {\it recfast.f} subroutine by {\it drecfast.f} one in available codes (CMBfast, CMBEASY etc.) for calculation of temperature and polarization $C_{\it l}$'s does not give correct estimation of possible changes to CMB power spectra caused by effect demonstrated here. It looks that collision term connected with Thomson scattering in electron density perturbed region and Boltzmann equation for photons must be generalized to take this effect into account properly. It will be the matter of the separate paper.

\section*{Conclusions}

In the field of adiabatic density perturbations the  rates of hydrogen and helium ionization-recombination processes slightly differ from 
those in the non-perturbed medium. On the one hand the rate of recombination increases due to somewhat  larger number density of plasma particles in the overdense region, on the other hand the photoionization is increased too due to positive temperature initial perturbation since number density of photons capable to ionize atoms is slightly 
higher, rising the  level of ionization. The  second effect prevails at the beginning of recombination epoch and competition of effects results into the positive additional overdensity of ions  and electrons. For initial perturbations with opposite sign (underdense region) all results are symmetric. The effect is prominent for large-scale initial perturbations ($k\le k^s_{dec}$) which never oscillate acoustically. Thus, the maximal amplitude of perturbations of proton and electron relative number density  during the recombination epoch is by factor of $\simeq 5$ higher than amplitude of $\delta_b$, baryon  density perturbations. For helium it is $\simeq 18$ times higher, but practically does not influence the amplitude of perturbations of  electron relative number density because of  low level of ionization at $z<1200$. For initial perturbations of scales inside acoustic horizon at decoupling epoch ($k\ge k^s_{dec}$) the deviations of ion number density perturbations from baryon total density ones vanish owing to the temperature and density oscillations of photon-baryon plasma during recombination. They are small also for isothermal initial perturbations. 

At lower redshifts, when ionization becomes residual and baryon mass density perturbation increases the electron number density perturbations  $\delta_e$  becomes   lesser than $\delta_b$  independently of scale of baryon density perturbation. The difference  acquires $\sim 20\%$ at $z\sim 200$.

Revealed deviations of electron number density perturbations amplitudes from baryon total density ones at cosmological recombination epoch 
do not lead to appreciable additional local temperature fluctuations in matter or thermal radiation. But they result in faint optical depth ''corrugation'' of last scattering surface: $\delta z_{dec}/(z_{dec}+1)\approx -0.33\delta_b$ at scales larger than sound horizon. Taking them into account may improve the  agreement
between theoretical predictions and observable data of current WMAP and future PLANCK missions on large-scale CMB anisotropy and polarization.

\section*{Acknowledgments}
I would like to thank Bohdan Hnatyk, Stepan Apunevych and Yurij Kulinich for useful discussions and comments. I am also grateful to Michal Ostrowsky for hospitality and possibility to work one month in Astronomical observatory of Jagellonian University, where the stimulating academic atmosphere allowed this work to progress. This work  is performed in the framework of  state project No 0104U002125 (Ministry of Education and Science of Ukraine).

\appendix

\section[]{The values of atomic constants and coefficients of approximation formulae used in numerical calculations}

\begin{table}
\caption{Atomic constants and coefficients of approximation formulae}
\label{tab}
\begin{center}
\begin{tabular}{c|c|c|c}
\hline 
& & &\\
Constant&Value&Source&Formula \\
& & &\\
\hline 
& & &\\
$\chi_{\rm HI}$&$2.17871122\cdot 10^{-18}$ J & \citet{seager1999}&(\ref{SahaHII}),(\ref{SahadxHII}) \\
$\chi_{\rm HeI}$&$3.9393393\cdot 10^{-18}$ J & \citet{seager1999}&(\ref{SahaHeII}), (\ref{SahadxHeII})\\
$\chi_{\rm HeII}$&$8.71869443\cdot 10^{-18}$ J & \citet{seager1999}&(\ref{SahaHeIII}),(\ref{SahadxHeIII}) \\
$h\nu_{\rm H2s}$&$1.63403509\cdot 10^{-18}$ J & \citet{seager1999}&(\ref{eqHII}),(\ref{eqdxHII}) \\
$h\nu_{ps}$&$3.30301387\cdot 10^{-18}$ J & \citet{seager1999}&(\ref{eqHeII}),(\ref{eqdxHeII}) \\
$h\nu_{\rm HeI2^1s}$&$9.64908312\cdot 10^{-20}$ J & \citet{seager1999}&(\ref{eqHeII}),(\ref{eqdxHeII}) \\
$h\nu_{2s-1s}$ (HI)&$5.4467613\cdot 10^{-19}$ J & \citet{pequignot1991}&(\ref{beta}) \\
$h\nu_{2s-1s}$ (HeII)&$6.36325429\cdot 10^{-19}$ J & \citet{hummer1998}&(\ref{beta}) \\
$\lambda_{\rm H2p}$&121.567 nm & \citet{seager1999,verner1996}&(\ref{eqHII}),(\ref{eqdxHII})\\
$\lambda_{\rm HeI2^1p}$&58.4334 nm & \citet{seager1999,verner1996}&(\ref{eqHeII}),(\ref{eqdxHeII})\\
F&1.14& \citet{seager1999}&(\ref{alphaHI}) \\
a&4.309& \citet{pequignot1991}&(\ref{alphaHI}) \\
b&-0.6166& \citet{pequignot1991}&(\ref{alphaHI}) \\
c&0.6703& \citet{pequignot1991}&(\ref{alphaHI}) \\
d&0.5300& \citet{pequignot1991}&(\ref{alphaHI}) \\
q&$1.80301774\cdot 10^{-17}$& \citet{hummer1998}&(\ref{alphaHeI}) \\
p&0.711& \citet{hummer1998}&(\ref{alphaHeI}) \\
$T_1$&$1.30016958\cdot 10^5$ K& \citet{hummer1998}&(\ref{alphaHeI}) \\
$T_2$& 3K& \citet{hummer1998}&(\ref{alphaHeI}) \\
$\Lambda_{\rm H}$&8.22458 $c^{-1}$&\citet{goldman1989}&(\ref{eqHII}),(\ref{eqdxHII}) \\
$\Lambda_{\rm He}$&51.3 $c^{-1}$&\citet{drake1969}&(\ref{eqHeII}),(\ref{eqdxHeII}) \\
& & &\\
\hline 
\end{tabular}
\end{center}
\vspace{0.3cm}
\end{table}

\end{document}